# Phase equilibria - thermal conductivity relationship within multicomponent Phase Change Materials from 273 K up to above the melting temperature

Anh Thu Phan, Aïmen E. Gheribi*and Patrice Chartrand

*CRCT - Center for Research in Computational Thermochemistry, Department of Chemical Eng., Polytechnique Montréal (Campus de Université de Montréal), Box 6079, Station Downtown, Montréal, Québec, Canada H3C 3A7*

## Abstract

Among all the properties required for the design of the next generation of PCM (density, heat capacity, thermal expansion, latent energy, volume change upon melting, corrosion rate, etc.) the thermal transport properties are by far the least known, especially for molten salt mixtures and solid solutions. We present in this paper a theoretical framework for accurate predictions of thermal conductivity of multicomponent salt-based PCM, from 273.15 K up to above melting temperature. The solid phase is considered as a microstructure with its proper temperature dependent parameters: phase volume fraction, grain size distribution, porosity, etc. As case studies, five new potential PCMs for CSP applications are considered. Their thermal conductivity is estimated as a function of temperature, from room temperature to 200 K above their melting point. The predictive capability of the proposed framework is discussed based on a comparison with available experimental data. The effect of equilibrium and non-equilibrium microstructural parameters (i.e. phase fraction, phase composition, average grain size, inter-grain, and intra-grain porosity) on the effective thermal conductivity of the solid states of the promising chloride PCMs is discussed. Lastly, recommendations for the design of next generations of PCM materials are suggested in order to improve their thermal transport properties.

*corresponding authors: email: aimen.gheribi@polymt.ca Tel: +1 514-340-4711 ext. 2316

# 1 Introduction

Sensible or latent heat materials are widely used to store thermal energy. Researchers are recently focusing on phase change materials (PCMs), which facilitate heat storage via their latent heat of fusion, as opposed to sensible heat storage materials having relatively low thermal capacity and requiring a considerable volume of the heat storage unit.[1-2] Consequently, there has been remarkable progress in the development of PCMs for practical applications.[2-5] Molten-salt energy storage (MSES) has been used commercially as a Thermal Energy Storage (TES) method to store heat collected by Concentrated Solar Power (CSP). The heat is then converted into superheated steam to run conventional steam turbines and produce electricity in inclement weather or at night. Various simple eutectic mixtures of different salts can be employed as PCMs for MSES. Chloride mixtures are widely used as PCMs due to their relatively high heat storage capacity and low cost (except those containing LiCl), despite their low thermal conductivity.[1] The design of new PCMs for the next generation of CSP is a real challenge which first of all requires to identify eutectics on multicomponent systems. In general, the identification of eutectics and therefore potential new PCMs is simply done by analysing available phase diagrams. For salts systems, most of the binary and several ternary phase diagrams are available, but very few diagrams for quaternary and higher order systems have been published, limiting considerably the exploration of PCMs offering better performances in terms of energy storage capacity and charging/discharging heat process. In principle, the FactSage software and FTsalts database[6] allows to determine potential PCMs within binary and ternary salts systems (chlorides, fluorides, bromides, iodides, carbonates, nitrates, nitrites, hydroxides, sulphates,…) by identifying eutectics or simple minima upon liquidus (azeotrope-like), via phase diagram calculations. The calculations of quaternary and higher order phase diagrams are however much more complex and not really accessible in the entire temperature range (above the liquidus temperature) and composition. To identify the local minima upon the liquidus surface, we have developed a software combining FactSage and the Mesh Adaptive Direct Search (MADS) algorithm,[7-10] devoted to solve "blackbox" minimization problems under non-smooth and non-linear constraints. The identification of a PCM consists in searching, via a particular minimization procedure based on MADS,[7] null values in the freezing range of multicomponent systems. By employing the procedure of Gheribi et al.[11], several potential low-cost PCMs (< 750 USD/metric) operating at a temperature of 663 K can be identified, allowing the generation of steam upon efficient heat transfer while restricting the corrosion rate of the containing materials. Of course, a phase change material suitable for CSP is not only a eutectic or a simple minimum upon the liquidus surface in a certain range of melting temperature. An optimal PCM must meet a certain number of drastic constraints and objectives on raw materials prices and long term availability, on thermophysical and thermal transport properties and corrosion resistance with respect to the low cost steel tanker as well. Due to the increasingly growing demand (mainly for battery applications) and a high cost (>7000 USD/ton), PCMs containing Li-based salts (LiCl, LiF, $Li_2CO_3$…) are not suitable for next generation of CSP. For the same reasons, on top of Li- based salts, the materials containing strategic metals, in particular rare earth element or those very scarce such as Cs and Rb based salts cannot be considered. Among the potential PCMs identified within multicomponent systems, only those with the best material performance can be retained as CSP materials. An optimal PCM must ideally meet the following requirements[12]:

- largest possible volumetric latent heat of fusion, > 450MJ/$m^3$, to maximize the thermal storage capacity; largest possible volumetric heat capacity, > 2MJ/($m^3$.K), to maximize the sensible heat;
- largest possible thermal conductivity in solid and liquid states, >1 W/(m.K), to provide a minimum temperature gradient and consequently accelerate the heat transfer processes;
- a small relative volume change upon melting, <15%, to maximize the tanker capacity;
- a minimal system vapor pressure at working temperature, <100 Pa;
- a minimum corrosion rate of the steel tanker, <50 μm/year.

Note that not only the PCM composition and melting temperature can be determined via FactSage but also the volumetric latent heat of fusion, the volumetric heat capacity, the volume change upon the melting. Even though the corrosion product can be also predicted with the FactSage software, the corrosion rate can only be determined experimentally, via a series of corrosion tests. Thermal conductivity or thermal diffusivity are key properties in the design of CSP plants as they are necessary for estimating heat transfer rates using fundamental heat transfer equations.[13] Unfortunately, a severe lack of experimental on thermal conductivity of molten salt mixtures and solid salt systems is observed.

For pure molten salt compounds, despite several experimental thermal conductivity datasets reported for many systems, only a few of them may be considered as reliable, mainly the most recent ones. In general, up to more than 100% discrepancy is observed between the different experimental datasets.[14] Experimental data are scarce for binary molten salts systems and almost inexistent for ternary and higher order systems. For simple ionic solid compounds (NaCl, KCl, LiF…), there are several consistent experimental datasets for the temperature dependent thermal conductivity (diffusivity) reported in the literature. However, solid systems rarely contain simple compounds, they are in general consisted of a phase assemblage of more complex compounds, resulted from a reaction between endmembers (e.g $NaCl + MnCl_2 = NaMnCl_3$), and solid solutions. The thermal conductivity (diffusivity) of most complex salts and solid solutions remains unknown from an experimental point of view. Note that the thermal transport properties of solid systems do not only depend on the phase amount, but on several microstructural parameters: phase's average grain size and phase's grain size distribution, microstructure configuration, porosity size and distribution, etc.

In recent years, we have developed theoretical approaches for accurately predicting the thermal conductivity (diffusivity) of molten salt mixtures and solid salt microstructures.[14-19] These approaches combine a series of theoretical models which can be parameterized either (i) via a CALPHAD-like method considering experimental data, not necessarily on thermal conductivity, but on other thermophysical properties (heat capacity, thermal expansion, elastic constants, velocity of sound,…), or (ii) ab initio via molecular dynamics or DFT simulations.

The aim of this paper is to propose a clear methodology for an accurate prediction of the thermal conductivity of PCMs from standard temperature (273.15 K) up to above the melting temperature. This methodology summarizes the different models for both liquid and solid states that were developed and for which the predictive capability has been validated in our prior works. As case studies, the thermal conductivity of 5 promising chloride based PCMs, melting around 390$^{o}$C, are predicted as a function of temperature. The variation of thermal conductivity with phase volume fraction and upon the solid-solids phase transition and melting is discussed. Lastly, the impact of microstructural parameters upon the thermal transport of "real" salts solid systems is quantified and discussed. Recommendations to improve the thermal transport properties of PCMs for the next generation CSP are proposed.

# 2 Methodology

The methodology to predict and represent the thermal conductivity of PCMs from standard temperature (273.15 K) up to hundreds K above the melting temperature consists in a series of theoretical models supported by DFT based atomistic simulations. These models were already published and will not be repeated in details here, but we wish to make a synthesis of the theoretical framework allowing to predict the thermal conductivity of PCMs in a wide range of temperatures. At the melting temperature, a PCM undergoes a first order phase transition, i.e from a solid to a liquid state. The solid state consists in a microstructure with intrinsic characteristic called microstructural parameters which are: phase distribution, phases' volume fractions, microstructure morphology, phases' average grain sizes, grain size distribution, porosity level, porosity size and distribution. The liquid phase consists, in principle, of a homogenous mixture with, depending on the system, different degrees of chemical ordering, from quasi fully dissociated mixtures to those with a strong short range ordering leading to the formation of coordination complexes and polymers. Note that, in principle for a PCM no liquid/solid biphasic region exists.

**Molten mixtures -** In a prior study,[14] based on the kinetic theory, we have proposed a theoretical model for the thermal conductivity of a molten salt compound, $\lambda_{lc}$, as a function of temperature. To achieve this, the Boltzmann transport equation has been solved assuming ions as hard spheres, i.e. the interaction potential is assumed to be null above a certain critical radius (hard-sphere radius) and infinite below this radius. This model is in principle, predictive as it contains no information on the thermal conductivity but only a few key thermodynamic properties. It is expressed as follows:

$$\lambda_{lc}(T) = 4.33\,\frac{C_{V,m}U_m}{3n_0V_m}(r_a + r_c)\left[1 - \alpha_m \cdot \left(\gamma + \frac{1}{3}\right)(T - T_m)\right] \quad (1)$$

Where $C_{V,m}$, $\alpha_m$, $U_m$, $V_m$ and $\gamma$ are respectively the heat capacity at constant volume, the thermal expansion coefficient, the velocity of sound, the molar volume and the Grüneisen constant all taken at melting temperature, $T_m$. $n_0$, $r_a$ and $r_c$ are the number of ions per formula, the effective anionic and cationic radius. According to this formalism, the temperature dependent thermal conductivity of molten salts can be predicted providing that $C_{V,m}$, $\alpha_m$, $U_m$, $V_m$ and $\gamma$ are known with an appreciable accuracy. Overall, the model has shown a very good predictive capability with most molten salts matching with most recent experimental data within a margin of less than 20 %, which is about the typical experimental error. Note that Eq. 1 has, in general, a better accuracy than molecular dynamics simulation (MDS) based on Fumi-Tosi potentials,[14] which includes the dipole(quadrupole)-dipole(quadrupole) polarization and dispersion contributions. Thereafter, given the severe lack of experimental data, a similar theoretical approach has been developed to describe the composition (X) dependence of the thermal conductivity of non-reciprocal molten salt mixtures, $\lambda_{ls}$, i.e mixtures with only one common anion ($Cl^-$, $F^-$, $OH^-$, $CO_3^{2-}$, $SO_4^{2-}$, $NO_3^-$,…). The proposed model for molten salt mixtures is also based on the kinetic theory and requires only few key thermodynamic data for its parameterization. In a nutshell, the composition dependence of the thermal conductivity of molten salts mixtures is expressed as:[20]:

$$\lambda_{ls}\,(\boldsymbol{X},T) = \lambda_\sigma(\boldsymbol{X},T)\left[1 - \delta_M^\lambda(\boldsymbol{X},T)\right] \tag{2}$$

where $\boldsymbol{X} = \{X_1, X_2 \ldots X_n\}$ denotes the composition vector and $\lambda_\sigma$ is the ideal thermal conductivity of a hypothetical molten mixture consisted of ions with identical centers of mass. From the kinetic theory, it can be shown that $\lambda_\sigma$ is approximated via a linear relationship with composition of the ratio: $\left[\left(C_{V,m} \cdot U_m\right)/(n_0 \cdot \lambda_{lc})\right]$.[20] Then, $\delta_M^\lambda$ is a function describing the mass fluctuation effect upon the thermal conductivity and therefore its composition dependence. We have shown[20] that $\delta_M^\lambda = 0.4872\,\lambda_\sigma \cdot g_{mass}/\left[k_B U_m (n_0 N_A/V_m)^{2/3}\right]$ with $g_{mass} = \sum_0^N X_i (M_i/M - 1)^2$ ($k_B$ and $N_A$ are Boltzmann's and Avogadro's constants respectively) is a good approximation for a large variety of molten salt mixtures. Note that as the thermal diffusivity, $a$, is defined from the thermal conductivity, $a = \lambda/(\rho.C_p)$, it can also be predicted from Eq. 2 providing temperature dependent heat capacity and density are known with a good precision. In CALPHAD like databases, in particular FTsalt, the critically assessed model parameters for both heat capacity and density are available for pure, binary and ternary systems. Then, considering a reliable extrapolation technique,[21-22] the density and heat capacity can be extrapolated to higher order systems to predict the thermal diffusivity of multicomponent molten PCMs.

**Solid state systems** – Predicting the thermal conductivity of solid state PCMs requires, first, an accurate knowledge of the phases constitutions and the phases volume fractions as a function of temperature. For solid systems, a phase may be either a stoichiometric compound or a solution. Note that, similarly to molten mixtures, a solid solution may contain a certain degree of short-range ordering. For each phase, in addition to the temperature dependent thermal conductivity, its volume fraction must also be determined as a function of temperature. Note that the evolution of the phase volume fraction is defined through its thermal expansion coefficient.

The representation and the prediction of the phase equilibria within multicomponent systems is nowadays mainly performed via the CALPHAD method. The CALPHAD (CALculation of PHAse diagrams) method is an effective computational approach to represent the thermodynamic properties of various phases through Gibbs energy models, and compute phase equilibria of multicomponent systems via Gibbs energy minimization under compositions constraints. To predict the phase equilibria within multicomponent systems, reliable databases must be available, containing critically assessed Gibbs energy parameters for at least unary, binary, and ternary subsystems. The prediction of phase equilibria within higher order systems is then performed based on suitable extrapolation of the phase's Gibbs energy.[21-22] For chloride multicomponent systems, the predictive capability of phase equilibria using the FTsalt database in the FactSage thermochemical Calculation Package[23] has been proven highly accurate.[24] In addition, all local minima upon liquidus surface of a multicomponent system can be identified by the FactOptimal software which is also included in FactSage[23] by minimization of the freezing range via the Mesh Adaptive Direct Search (MADS) algorithm.[7-8] As an example, the invariant points of the quaternary and quinary subsystems within LiF-LiCl-NaF-NaCl-KF-KCl-$SrF_2$-$SrCl_2$ were predicted with an appreciable accuracy,[24] showing difference of less than 20 K for the liquidus temperature and 0.025 mole fraction for the composition in comparison with the available experimental data.[25] When the phase distributions are quantified, a temperature and/or composition dependent thermal conductivity must be affected to each phase. The thermal conductivity of electrically insulating stoichiometric compounds is mainly of vibrational origin (phonon) and, depending on optical properties (surface emissivity), may also contain a significant radiative contribution at higher temperature ($T \gtrsim 1000\ K$). The vibrational (phonon) thermal conductivity of solid salt compounds, $\lambda_{sc}$, is, in general, accurately represented by the following expression:[16, 26]

$$\lambda_{sc} = A \cdot \frac{\bar{M}\theta_D^3\delta}{\gamma^2 T n^{2/3}} \cdot exp\left(\frac{\theta_D}{3T}\right) \quad (3)$$

Where $\bar{M}$ is the average atomic mass; $\delta^3$ is the volume per atom; $\theta_D$ is the Debye temperature; $n$ is the number of atoms per primitive cell; and the constant $A = (2.43 \cdot 10^{-8})/[1 - 0.514/\gamma + 0.228/\gamma^2]$ is a material constant which only depend on the Grüneisen parameter, $\gamma$.[26]

Due to the disorder scattering, the thermal conductivity of solid solutions, $\lambda_{ss}$, is, in general, much lower than that of the corresponding compounds. According to the kinetic theory, the thermal conductivity of solid solutions may be directly expressed through the thermal conductivity of the parent materials, $\lambda_{ss}^{id}$, and a function, $\boldsymbol{u}$ (which is temperature and composition dependent), which accounts for a disorder scattering as:[16, 27]

$$\lambda_{ss}(T,\boldsymbol{X}) = \lambda_{ss}^{id}(T,\boldsymbol{X}) \cdot \frac{arctan\,(\boldsymbol{u})}{\boldsymbol{u}} \qquad (4)$$

Note that $\lambda_{ss}^{id}$ is quantified from the physical properties of the solid solution using Eq. 3, in particular the composition dependent atomic volume, Debye temperature and the Grüneisen parameter. In our prior work,[16] starting from Abeles's formulation,[27] we have proposed an original formulation for $\boldsymbol{u}$ which has the particularity to be directly linked to $\lambda_{ss}^{id}$:

$$\boldsymbol{u}(T,\boldsymbol{X}) = \sqrt{\frac{\pi\theta_D(\boldsymbol{X})\bar{M}(\boldsymbol{X})}{2\hbar N_A B_s(T,\boldsymbol{X})} \cdot \lambda_{ss}^{id}(T,\boldsymbol{X}) \cdot \mathit{\Gamma}(T,\boldsymbol{X})} \qquad (5)$$

where $\hbar$ and $B_s$ are respectively the Planck's constant, and the isentropic bulk modulus of the solid solution. $\mathit{\Gamma}$ is the disorder scattering parameter resulting from the mass fluctuation ($\mathit{\Gamma}_M$) and the local strain field fluctuation ($\mathit{\Gamma}_S$) which are assumed to be independent: $\mathit{\Gamma} = \mathit{\Gamma}_M + \mathit{\Gamma}_S$.[16] The mass fluctuation term depends on composition only: $\mathit{\Gamma}_M(\boldsymbol{X}) = \sum_i X_i[M_i/M(\boldsymbol{X}) - 1]^2$ ($M_i$ and $M$ are the molecular weight of each constituent of the solution and the average molecular weight) while the local stain field fluctuation is, in principle, a function of the lattice constants and all the elastic constants of the crystal and therefore depends on both temperature and composition. Abeles[27] proposed the following expression to describe strain field contribution to the disorder scattering parameter: $\mathit{\Gamma}_S(\boldsymbol{X},T) = \varepsilon \sum_i X_i[\delta_i(T)/\delta(\boldsymbol{X},T) - 1]^2$ which is a simplification as it does not explicitly consider the elastic constants of the crystals. $\varepsilon$ is an empirical parameter depending mainly on the nature and the magnitude of chemical binding within the solid solution. For chloride solid solutions, we have shown[16] that $\varepsilon = 5$. Note that, for salt solid solutions, the composition dependence of $\delta$ can be accurately approximated via the Vegard rule while that of $\theta_D$ and $\gamma$ can be estimated via the well-known Kopp-Neuman approximation. The composition dependence of adiabatic bulk modulus of salt solid solutions can be estimated from the Flancher and Barsh model.[28] In other words, according to the present formalism the composition dependence of the thermal conductivity of a simple solid solution can be estimated from thermal conductivity parameters of end-members (compounds).

**Prediction of model parameters for complexes compounds** – For simple alkali compounds (NaCl, KCl, $MgCl_2$…), there are in general several consistent sets of experimental thermal diffusivity data available in the literature. However, a severe lack of experimental data is observed for transition metal salts ($MnCl_2$, $NiCl_2$, $CuCl_2$…) and for binary and higher order alkali compounds (e.g. $K_2CoCl_4$ and $KCaCl_3$). To alleviate the lack of experimental data, we have developed in our prior work[15-16] a CALPHAD-like method to predict the thermal conductivity parameters from experimental information on heat capacity, thermal expansion and bulk modulus as a function of temperature. In a nutshell, according to this methodology, the high temperature limit of the Debye temperature and Grüneisen parameter involved in the temperature dependent expression of the thermal conductivity (Eq. 3) are determined via a multi-objective fitting procedure based on the Debye-Grüneisen formalism to reproduce available experimental data on thermal expansion, heat capacity and the elastic constants from low temperature up to high temperature (ideally from 0 K up to the melting temperature). When experimental data are not available, both the Debye temperature and Grüneisen parameter are calculated by *ab initio* along with thermodynamic and volumetric properties by combining Density Functional Theory (DFT) with the Quasiharmonic Approximation (QHA) method. However, QHA is known to overestimate the lattice expansion contribution upon the thermodynamics properties leading to the violation of the Maxwell relations and therefore to the consistency between the thermodynamics properties.[29-30] To overcome these limitations, we have developed an extension of the QHA approach (thermodynamically self-consistent (TSC) method [29-31]) which ensures the respect of the Maxwell relations.[29-31] The TSC method has shown a good predictive capability for a large variety of materials not only for thermodynamic properties[29-30] but also for thermal conductivity through its accurate prediction of both Debye temperature and Grüneisen parameters. To illustrate the predictive capability of TSC method, the calculated thermal expansion (defining the temperature dependence of the volume fraction) and the thermal conductivity of CsCl single crystal (cubic-B2, Pm-3m) are shown in Figure 1 in comparison with available experimental data.

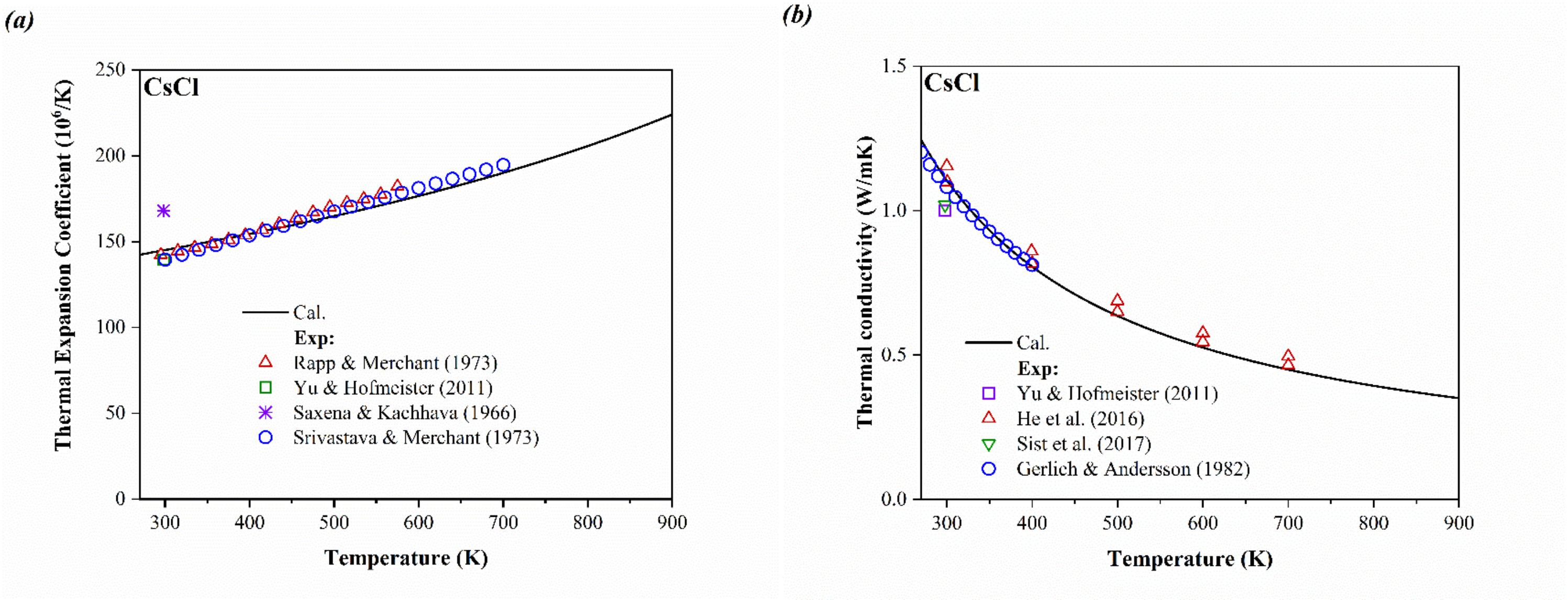

*Figure 1 Calculated (a) thermal expansion coefficient and (b) thermal conductivity of B2-CsCl via TSC method in comparison with available experimental data.*[32-38]

The agreement is found to be very good in the entire range of temperature. Such good agreement originates, however, from the fact that CsCl crystalizes in a simple structure and it is almost isotropic with respect to its elastic properties. For complex compounds (such as $Na_3AlF_6$, $Na_5Al_3F_{14}$ or $Na_4Ca_4Al_7F_{33}$) the capability of TSC to predict the thermal expansion and thermal conductivity is found to be also good, in general less than 20 % of error is observed.[17-19, 31]

**Thermal conductivity of microstructures** – The solid state can be effectively described as a given phase assemblage with intrinsic parameters called microstructural parameters. In our approach, the microstructural parameters are:

**A- Equilibrium properties obtained via FactSage from PCM nominal composition**

(i) Phase's (compounds and/or solutions) amounts (mole or weight fraction) as a function of temperature (Note that due to chemical reactions, phases may appear and/or disappear as temperature changes. The phase constitutions of a solid state PCM may be completely different between 273.15 K and melting temperature);

(ii) Phase's temperature dependent volume fraction determined from its density and thermal expansion coefficient;

**B- Non-equilibrium parameters determined via characterization experiments**

(i) Microstructure morphology (e.g. uniformly distributed grains, precipitate in a primary phase matrix, dendritic precipitate, etc.);

(ii) Average grain size of all phases of the system, $\langle d_i \rangle$, and their standard deviations;

(iii) Crystallographic grain characteristics (shape, grain-grain coordination, grain boundaries chemistry, etc.) and identification of phases of the matrix, if present;

(iv) Intra-grain and inter-grain porosity level and average size (Intra-grain porosity is the nano-porosity within a grain while inter-grain porosity is the micro to macro-porosity between grains);

(v) Porosities distribution.

The microstructural parameters related to the equilibrium properties (A) can be accurately determined via the FactSage software, providing both thermodynamic and physical properties, and databases containing sufficient information on unary, binary and ternary systems for a reliable extrapolation for higher order systems. There are several approaches to describe the impact of the phase assemblage upon the thermal conductivity of the microstructure. The phase assemblage model depends mainly on the microstructure configuration and the porosity distribution. In its solid-state, a salt system is consisted, in general, of grains of almost uniformly distributed porosity even though they may have very different grain sizes. In this case, the effective medium theory (EMT) can, in general, better describe the phase assemblage effect upon the thermal transport. According to the EMT approach the thermal conductivity of a solid system with zero inter-grain porosity (macro-porosity), $\lambda_{eff}^0$, is calculated by solving:[39-40]

$$\sum_{i=1}^{n} \varphi_i(T,\boldsymbol{X}) \cdot \frac{\lambda_i(T,\langle d_i \rangle, p_i^{intra}) - \lambda_{eff}^0(T,\boldsymbol{X},\langle \boldsymbol{D} \rangle, \langle \boldsymbol{P^{intra}} \rangle)}{\lambda_i(T,\langle d_i \rangle, p_i^{intra}) + \lambda_{eff}^0(T,\boldsymbol{X},\langle \boldsymbol{D} \rangle, \langle \boldsymbol{P^{intra}} \rangle)} = 0 \qquad (6)$$

Where $\varphi_i$, $\lambda_i$, $\langle d_i \rangle$ and $p_i^{intra}$ denotes respectively the volume fraction, thermal conductivity, the average grain size and the intra-grain porosity (nano-porosity) also of the phase $i$. $\langle \boldsymbol{D} \rangle = \{\langle d_1 \rangle, \langle d_2 \rangle \ldots \langle d_N \rangle\}$ and $\langle \boldsymbol{P^{intra}} \rangle = \{p_1^{intra}, p_2^{intra} \ldots, p_N^{intra}\}$ are the phase's average grain size and intra-porosity vectors.

The following expression have been derived from the Boltzmann thermal equation (BTE) to consider the grain size dependence upon the thermal conductivity of individual grains:[41]

$$\lambda_i(T,d) = \lambda_i(T,d_i \to \infty) \cdot \left[1 - \sqrt{\frac{\sigma_i(T)}{< d_i >}} arctan \sqrt{\frac{< d_i >}{\sigma_i(T)(T)}}\right] \quad (7)$$

Where $\sigma_i$ is a temperature dependent phonon characteristic length within the grain and it can be estimated from the Debye temperature, the thermal conductivity of single crystal, the thermal expansion coefficient, heat capacity and the Grüneisen parameter (see ref.[41] for more details). Note that the grain size dispersion can be considered by weighting the grain size via Gaussian damping factor depending on the standard deviation of the grains size distribution. From an empirical point of view, it is known that up to a certain level, the intra-porosity (nano-porosity) decreases the thermal conductivity of grains almost linearly, but above a certain porosity level (in general, depending on the materials, it is around 10-20%) the intra-porosity decreases the thermal conductivity more rapidly. To include this behaviour, Odelevskii[42-43] proposed the following expression:

$$\lambda_i(T, \langle d_i \rangle, p_i^{intra}) = \lambda_i(T, \langle d_i \rangle, p_i^{intra} = 0) \cdot \frac{1 - p_i^{intra}}{1 + m \cdot p_i^{intra}} \quad (8)$$

In which *m* in an empirical parameter related to the pore size and distribution, in general for insulating material it varies between 5 and 10. Note that at low porosity the above formalism leads to a linear behaviour of the thermal conductivity of grains with intra-porosity. *m* can also be linked to the porosity level above which the thermal conductivity starts to decrease more rapidly than (1-$p_i^{intra}$). In our approach, the inter-grain porosity (macro-porosity, $P^{inter}$) effect upon the thermal conductivity of a solid system is described effectively and globally, i.e. the inter-grain porosity is considered as an average parameter for the entire system (no inter-grain-grain porosities are considered). Without a loss of generality, the thermal conductivity of a solid system, $\lambda_{eff}^{ms}$, can be expressed as:

$$\lambda_{eff}^{ms}(T, X, \langle \boldsymbol{D} \rangle, \langle \boldsymbol{P^{intra}} \rangle, P^{inter}) = \lambda_{eff}^{0}(T, X, \langle \boldsymbol{D} \rangle, \langle \boldsymbol{P^{intra}} \rangle) \cdot \tau(P^{inter}) \quad (9)$$

Where $\tau$ is a function describing the average inter-grain porosity level upon the entire solid state system. We have shown that in the case of inter-grain porosity, there exists a critical porosity, $P_C$, beyond which the thermal conductivity starts to decrease rapidly.[44-46] This rapid decrease is interpreted as a second order phase transition associated to grains-porosities symmetry breaking. In a nutshell, according to this theory, when the inter-grain porosity increases, the probability of grain-grain contact decreases dramatically, leading to a critical decrease of the phonon mean free path and therefore to the thermal conductivity as well. Note that another interpretation of this critical behavior can be found in the percolation theory.[47] This critical behavior has been formulated as:[44]

$$\tau\left(P^{inter}\right) = \left[1 + \int_{0}^{P^{inter}} \chi\left(P^{inter}\right) dP^{inter}\right] \tag{10}$$

Where $\chi$ is a critical function depending on the critical porosity as:[44]

$$\chi\left(P^{inter}\right) = A_{\pm} ln \left[\frac{1 + \left(P^{inter}/P_C\right)^{\pm 3}}{1 + (P^{inter}/P_C)^{\pm 3}}\right] \tag{11}$$

# 3 Results and discussions

Considering promising PCMs which should (i) have low cost; (ii) not react with the tanker's material; (iii) be available in large quantity; and (iv) have a melting temperature of 390 ± 10°C (or 663 ± 10 K), in our previous study, we have shown that chlorides are the most suitable systems.[11] Beside Li-, Cs- and Rb-based salts, rejected because of their high prices, fluorides and hydroxides are also unsuitable because of either their high melting temperature range or their ability to corrode the low-cost steel tanker. Thereafter, we have identified five low-cost PCMs with the FactOptimal software (Table 1), and we have calculated the thermal conductivity of these PCMs from the standard temperature of 273.15 K up to 200 K above the liquidus temperature. Among the five PCMs of interest, there is 1 ternary system (E1), 3 quaternary systems (E2, E3, and E4), and 1 quinary system (E5). In most works presenting potential PCMs, only ternary multi-component PCMs are investigated. The fact that we can identify and do calculations for multi-component systems is very important in fine-tuning thermal physical properties and designing the most suitable PCMs for TES applications. The nominal compositions of the 5 eutectic PCMs were calculated automatically by the robust, rapid, and reliable FactOptimal software and the FTsalt database.[23] Besides the occurrence of many compounds and solid solutions precipitated at the eutectic temperatures, several solid-state first-order phase transformations are observed from room temperature up to the melting temperature (Table 1).

*Table 1: Calculated eutectic compositions of multicomponent chloride systems namely E1, E2, E3, E4, and E5 as promising PCMs and the associated invariant reactions including phase fractions and phase compositions by using the FTsalt database and the CALPHAD method available in the Factsage thermochemical Calculation Software Package[23] ((*)Note that when a Rocksalt solid solution de-mixes forming a miscibility gap, Rocksalt#1 and Rocksalt#2 are the two solid solutions of the miscibility gap):*

| PCM | Phase | Composition | Space group |
|---|---|---|---|
| **E1** *(0.4870 KCl - 0.0324 $CaCl_2$ - 0.4806 $PbCl_2$)*<br>• Liquidus temperature of $T_{E1} = 672.3\ K$:<br>Liquid → 0.03240 $KCaCl_3$(s2) + 0.16887 $KPb_2Cl_5$ + 0.14287 $K_2PbCl_4$<br>• 1st Solid phase transformation at $T_{E1}^{S1} = 540.6\ K$:<br>0.03240 $KCaCl_3$(s2) + 0.16887 $KPb_2Cl_5$ + 0.14287 $K_2PbCl_4$ → 0.03240 $KCaCl_3$(s2) + 0.24030 $KPb_2Cl_5$ + 0.21430 KCl<br>• 2nd Solid phase transformation at $T_{E1}^{S2} = 400.0\ K$:<br>0.03240 $KCaCl_3$(s2) + 0.24030 $KPb_2Cl_5$ + 0.21430 KCl<br>→ 0.03240 $KCaCl_3$(s1) + 0.24030 $KPb_2Cl_5$ +0.21430 KCl | Liquid | 0.4870 KCl - 0.0324 $CaCl_2$ - 0.4806 $PbCl_2$ | |
| | Compound | $K_2PbCl_4$ | $I\bar{4}3d$ |
| | | KCl | $Fm\bar{3}m$ |
| | | $KCaCl_3$(s1) | *Pnma* |
| | | $KCaCl_3$(s2) | *I4/mcm* |
| | | $KPb_2Cl_5$ | *$P2_1/c$* |
| **E2** *(0.190 NaCl - 0.527 KCl - 0.260 $MgCl_2$ - 0.023 $CaCl_2$)*<br>• Liquidus temperature of $T_{E2} = 656.7\ K$:(*)<br>Liquid → 0.26000 Sln1 + 0.02263 $KCaCl_3$(s2) + 0.1 Rocksalt#1 + 0.07455 Rocksalt#2<br>• Solid phase transformation at $T_{E2}^{S1} = 408.3\ K$:<br>0.26000 Sln1 + 0.02288 $KCaCl_3$(s2) + 0.13140 Rocksalt#1 + 0.04300 Rocksalt#2<br>→ 0.26000 Sln1 + 0.02310 Sln2 + 0.13140 Rocksalt#1 + 0.0430 Rocksalt#2 | Liquid | 0.190 NaCl - 0.527 KCl - 0.260 $MgCl_2$ - 0.023 $CaCl_2$ | |
| | Sln1 (solid solution) | 0.200 $Na_2MgCl_4$ – 0.800 $K_2MgCl_4$ (at $T_{E2} = 656.7\ K$ & $T_{E2}^{S1} = 408.3\ K$)<br>0.113 $Na_2MgCl_4$ – 0.887 $K_2MgCl_4$ (at $T_{E2}^{S1} = 408.3\ K$) | *I4/mmm* |
| | Rocksalt#1 (solid solution) | 0.1817 NaCl – 0.8183 KCl (at $T_{E2} = 656.7\ K$ & $T_{E2}^{S1} = 408.3\ K$)<br>0.99263 NaCl – 0.00644 KCl – 0.00093 $CaCl_2$ (at $T_{E2}^{S1} = 408.3\ K$) | $Fm\bar{3}m$ |
| | Rocksalt#2 (solid solution) | 0.0181 NaCl – 0.9819 KCl (at $T_{E2} = 656.7\ K$)<br>0.91285 NaCl – 0.08219 KCl – 0.00496 $CaCl_2$ (at $T_{E2}^{S1} = 408.3\ K$) | $Fm\bar{3}m$ |
| | Sln2 (solid solution) | 0.010 $KMgCl_3$ - 0.990 $KCaCl_3$ | *Pnma* |
| | Compound | $KCaCl_3$(s2) | *I4/mcm* |
| **E3** *(0.4768 NaCl - 0.0700 $MgCl_2$ - 0.1400 $CaCl_2$ - 0.3132 $MnCl_2$)*<br>• Liquidus temperature of $T_{E3} = 664.6\ K$:<br>Liquid → 0.11413 Ilmenite + 0.01326 $Na_9Mn_{11}Cl_{31}$ + 0.14157 Rutile + 0.12168 Sln3<br>• 1st Solid phase transformation at $T_{E3}^{S1} = 652.8\ K$:<br>0.12036 Ilmenite + 0.01281 $Na_9Mn_{11}Cl_{31}$ + 0.14138 Rutile + 0.12059 Sln3 → 0.11257 Ilmenite + 0.04357 $Na_2Mn_3Cl_8$ + 0.14138 Rutile + 0.13855 Sln3<br>• 2nd Solid phase transformation at $T_{E3}^{S2} = 452.6\ K$:<br>0.16522 Ilmenite + 0.03106 $Na_2Mn_3Cl_8$ + 0.14008 Rutile + 0.1247 Sln3 → 0.16522 Ilmenite + 0.03106 $Na_2Mn_3Cl_8$ + 0.1247 Sln3 + 0.14008 Hydrophilite | Liquid | 0.4768 NaCl - 0.0700 $MgCl_2$ - 0.1400 $CaCl_2$ - 0.3132 $MnCl_2$ | |
| | Ilmenite (solid solution) | 0.4046 $NaMgCl_3$ – 0.5954 $NaMnCl_3$ (at $T_{E3} = 664.6\ K$)<br>0.3974 $NaMgCl_3$ – 0.6026 $NaMnCl_3$ (at $T_{E3}^{S1} = 652.8\ K$)<br>0.3585 $NaMgCl_3$ – 0.6415 $NaMnCl_3$ (at $T_{E3}^{S2} = 452.6\ K$) | $R\bar{3}c$ |
| | Rutile (solid solution) | 0.98894 $CaCl_2$ – 0.01106 $MgCl_2$ (at $T_{E3} = 664.6\ K$)<br>0.9902 $CaCl_2$ – 0.0098 $MgCl_2$ (at $T_{E3}^{S1} = 652.8\ K$)<br>0.99944 $CaCl_2$ – 0.00056 $MgCl_2$ (at $T_{E3}^{S2} = 452.6\ K$) | *P4₂/mnm* |
| | Sln3 (solid solution) | 0.1829 $Na_2MgCl_4$ – 0.8171 $Na_2MnCl_4$ (at $T_{E3} = 664.6\ K$)<br>0.1724 $Na_2MgCl_4$ – 0.8276 $Na_2MnCl_4$ (at $T_{E3}^{S1} = 652.8\ K$)<br>0.0856 $Na_2MgCl_4$ – 0.9144 $Na_2MnCl_4$ (at $T_{E3}^{S2} = 452.6\ K$) | *Pbam* |
| | Hydrophilite (solid solution) | 0.99944 $CaCl_2$ – 0.00056 $MgCl_2$ (at $T_{E3}^{S2} = 452.6\ K$) | *Pnnm* |
| | Compound | $Na_2Mn_3Cl_8$ | $R\bar{3}m$ |
| | | $Na_9Mn_{11}Cl_{31}$ | $R\bar{3}c$ |
| **E4** *(0.08694 NaCl - 0.62055 KCl - 0.01878 $CaCl_2$ - 0.27373 $CoCl_2$)*<br>• Liquidus temperature of $T_{E4} = 655.3\ K$: (*)<br>Liquid → 0.01837 $KCaCl_3$(s2) + 0.27373 $K_2CoCl_4$ + 0.08361 Rocksalt#1 + 0.05847 Rocksalt#2 | Liquid | 0.08694 NaCl - 0.62055 KCl - 0.01878 $CaCl_2$ - 0.27373 $CoCl_2$ | |
| | Rocksalt#1 (solid solution) | 0.9130 NaCl – 0.0820 KCl – 0.0050 $CaCl_2$ (at $T_{E4} = 655.3\ K$)<br>0.9935 NaCl – 0.0057 KCl – 0.0008 $CaCl_2$ (at $T_{E4}^{S1} = 400.0\ K$) | $Fm\bar{3}m$ |
| • Solid phase transformation at $T_{E4}^{S1} = 400.0\ K$:<br>0.01871 $KCaCl_3$(s2) + 0.27373 $K_2CoCl_4$ + 0.08662 Rocksalt#1 + 0.05478 Rocksalt#2 → 0.01871 $KCaCl_3$(s1) + 0.27373 $K_2CoCl_4$ + 0.08662 Rocksalt#1 + 0.05478 Rocksalt#2 | Rocksalt#2 (solid solution) | 0.1814 NaCl – 0.8186 KCl (at $T_{E4} = 655.3\ K$)<br>0.0162 NaCl – 0.9838 KCl (at $T_{E4}^{S1} = 400.0\ K$) | $Fm\bar{3}m$ |
| | Compound | $K_2CoCl_4$ | *Pnma* |
| | | $KCaCl_3$(s1) | *Pnma* |
| | | $KCaCl_3$(s2) | *I4/mcm* |
| **E5** *(0.47550 NaCl - 0.06776 $MgCl_2$ - 0.13911 $CaCl_2$ - 0.30913 $MnCl_2$ - 0.00850 $NiCl_2$)*<br>• Liquidus temperature of $T_{E5} = 663.1\ K$:<br>Liquid → 0.01287 $Na_9Mn_{11}Cl_{31}$ + 0.00780 Sln4 + 0.10927 Ilmenite + 0.14063 Rutile + 0.12518 Sln5<br>• 1st Solid phase transformation at $T_{E5}^{S1} = 652.9\ K$:<br>0.01248 $Na_9Mn_{11}Cl_{31}$ + 0.00784 Sln4 + 0.11457 Ilmenite + 0.14047 Rutile + 0.12428 Sln5 → 0.04248 $Na_2Mn_3Cl_8$ + 0.00782 Sln4 + 0.10700 Ilmenite + 0.14047 Rutile + 0.14177 Sln5<br>• 2nd Solid phase transformation at $T_{E5}^{S2} = 452.6\ K$:<br>0.03002 $Na_9Mn_{11}Cl_{31}$ + 0.00840 Sln4 + 0.15823 Ilmenite + 0.13919 Rutile + 0.12862 Sln5 → 0.03002 $Na_2Mn_3Cl_8$ + 0.00840 Sln4 + 0.15823 Ilmenite + 0.13919 Hydrophilite + 0.12861 Sln5 | Liquid | 0.47550 NaCl - 0.06776 $MgCl_2$ - 0.13911 $CaCl_2$ - 0.30913 $MnCl_2$ - 0.00850 $NiCl_2$ | |
| | Sln4 (solid solution) | 0.00002 $CaCl_2$ – 0.00221 $MgCl_2$ – 0.00853 $MnCl_2$ – 0.98924 $NiCl_2$ (at $T_{E5} = 663.1\ K$)<br>0.0020 $MgCl_2$ – 0.0080 $MnCl_2$ – 0.9900 $NiCl_2$ (T=652.9K)<br>0.00014 $MgCl_2$ – 0.00085 $MnCl_2$ – 0.99901 $NiCl_2$ ($T_{E5}^{S2} = 452.6\ K$) | $R\bar{3}m$ |
| | Ilmenite (solid solution) | 0.4003 $NaMgCl_3$ – 0.5958 $NaMnCl_3$ – 0.0039 $NaNiCl_3$ (at $T_{E5} = 663.1\ K$)<br>0.3942 $NaMgCl_3$ – 0.6022 $NaMnCl_3$ - 0.0004 $NaNiCl_3$ (at $T_{E5}^{S1} = 652.9\ K$)<br>0.3582 $NaMgCl_3$ – 0.6414 $NaMnCl_3$ – 0.0004 $NaNiCl_3$ ($T_{E5}^{S2} = 452.6\ K$) | $R\bar{3}c$ |
| | Rutile (solid solution) | 0.9892 $CaCl_2$ – 0.0108 $MgCl_2$ (at $T_{E5} = 663.1\ K$)<br>0.9903 $CaCl_2$ – 0.0097 $MgCl_2$ (at $T_{E5}^{S1} = 652.9\ K$)<br>0.9994 $CaCl_2$ – 0.0006 $MgCl_2$ ($T_{E5}^{S2} = 452.6\ K$) | $P4_2/mnm$ |
| | Sln5 (solid solution) | 0.1796 $Na_2MgCl_4$ – 0.8175 $Na_2MnCl_4$ – 0.0029 $Na_2NiCl_4$ (at $T_{E5} = 663.1\ K$)<br>0.1707 $Na_2MgCl_4$ – 0.8267 $Na_2MnCl_4$ – 0.0026 $Na_2NiCl_4$ (at $T_{E5}^{S1} = 652.9\ K$)<br>0.0855 $Na_2MgCl_4$ – 0.9141 $Na_2MnCl_4$ – 0.0003 $Na_2NiCl_4$ ($T_{E5}^{S2} = 452.6\ K$) | *Pbam* |
| | Hydrophilite (solid solution) | 0.9994 $CaCl_2$ – 0.0006 $MgCl_2$ ($T_{E5}^{S2} = 452.6\ K$) | *Pnnm* |
| | Compound | $Na_2Mn_3Cl_8$ | $R\bar{3}m$ |
| | | $Na_9Mn_{11}Cl_{31}$ | $R\bar{3}c$ |

According to Table 1, upon temperature decreases, phase transformations of the five investigated PCMs involve a lot of solid solutions, complex compounds, and liquid phases. When considering the evolution of the thermal conductivity of the PCMs with temperature, first and foremost, we have estimated the volumetric thermal expansion coefficient and thermal conductivity of liquid compounds as a function of temperature using Eq. 1 (Table A1 and Table A2). For common molten salts such as NaCl, KCl, $CaCl_2$, the thermal conductivity was estimated using Eq. 1 in our previous study[14] and the experimental physical properties, in particular the available experimental sound velocity.[48] Note that for molten $MgCl_2$, rather than using the available experimental data of sound velocity,[49] we have re-evaluated the sound velocity using *Ab Initial* Molecular Dynamic simulation (AIMD) giving a more consistent value with other experimental sound velocities of other alkali halides. The temperature dependent thermal conductivity calculated using Eq. 1 is in agreement with the results obtained via First Principle Equilibrium Molecular Dynamic simulation (FP-EMD). Our calculated thermal conductivity of molten $MgCl_2$ is then used to predict the temperature dependent thermal conductivity of KCl-$MgCl_2$ binary and NaCl-KCl-$MgCl_2$ ternary which is in good agreement with the available experimental data.[50] There is a lack of thermal conductivity data on molten $PbCl_2$, $MnCl_2$, $CoCl_2$, and $NiCl_2$. For those chlorides, their thermal conductivity is calculated from Eq 1. (Table A2) using the physical properties derived from the available experimental density of molten salts as a function of temperature[51] (Table A1) and the approximation $\gamma = 1$. In addition, all the solid compounds and end-members of solid solutions involved in the phase transformations of PCMs (Table 1) have been considered for estimating their physical properties, in particular, thermal conductivity (Tables A3 and A4). Parameters for some common chlorides (NaCl, KCl, $CaCl_2$, etc.), for which physical properties are widely available in the literature are easy to obtain. Other chlorides, especially complex solid compounds such as $KPb_2Cl_5$, $Na_9Mn_{11}Cl_{31}$, $NaMgCl_3$, $K_2CoCl_4$, $Na_2NiCl_4$, etc. are parameterized by using the TSC method, providing reliable physical properties, especially thermal conductivity (Tables A3 and A4).

Since the thermophysical properties of all the solid and liquid compounds involved in the phase transformations of the five promising PCMs are estimated with reliability and models for calculating the thermophysical properties of solid solutions, solid phase assemblages and liquid solutions are available, the evolution of thermophysical properties of the PCMs from 273.15 K up to 200 K above the melting point is computed subsequently. Except for the thermal conductivity, the evolution of all other physical properties of PCMs with temperature is shown by using the FactSage thermochemical Calculation Package[23] (Table 2). The larger the value of latent heat, the more favorable the material for TES. Among the five PCMs of interest, E5 has the largest value of latent heat and it also has the smallest volume change during liquid-solid state transformation at the eutectic temperature, allowing the best usage of the tank capacity (Table 2 – light green cells). A material with a large molar volume change like E1, E2 is unfavorable as potential PCMs (Table 2 – dark orange cells). Considering the investigated PCMs, E3 and E5 are the two PCMs satisfying some of the requirements for an optimal PCM[12] such as large volumetric latent heat of fusion, large volumetric heat capacity, and relatively small volume change upon melting (Table 2).

*Table 2: Calculated eutectic temperature ($T_E$), heat of fusion ($\Delta H^f$) and other physical properties including heat capacity ($C_p$), density ($\rho$), volume ($V$), thermal diffusivity ($\alpha_D$), and thermal conductivity ($\kappa$) of both solid assemblage and liquid phase at $T_E$ of the five potential PCMs (Note that results in this table were calculated for 1 mole of each PCM, favorable values are shown in light green cells and unfavorable values in dark orange cells):*

| Properties | E1 | E2 | E3 | E4 | E5 |
|---|---|---|---|---|---|
| $T_E$(K) | 672.3 | 656.7 | 664.6 | 655.26 | 663.1 |
| $\Delta H^f$ ($MJ/m^3$) | 294.86 | 346.75 | 498.48 | 243.71 | **561.36** |
| $C_{p,solid}$ ($MJm^{-3}K^{-1}$) at $T_E$ | 1.84 | 1.76 | 1.85 | 1.68 | **1.88** |
| $C_{p,liquid}$ ($MJm^{-3}K^{-1}$) at $T_E$ | 1.89 | 1.74 | 1.94 | 1.78 | **2.01** |
| $\rho_{solid}$ ($g/cm^3$) at $T_E$ | **4.36** | 2.03 | 2.36 | 2.31 | 2.39 |
| $\rho_{liquid}$ ($g/cm^3$) at $T_E$ | **3.57** | 1.73 | 2.11 | 2.05 | 2.26 |
| $\Delta V$ at $T_E$ | 22.05% | 17.50% | 11.88% | 12.35% | **5.70%** |
| $\kappa_{solid}$($10^{-2}Wm^{-1}K^{-1}$) at $T_E$ | 21.77 | **67.89** | 59.83 | 57.85 | 62.61 |
| $\kappa_{liquid}$($10^{-2}Wm^{-1}K^{-1}$) at $T_E$ | 23.06 | 48.41 | **53.08** | 41.43 | 50.42 |

After calculating physical properties data with FactSage, calculations of thermal conductivity can be performed. By solving Eq. 6 using the volumetric fraction of phases obtained from the FTsalt database and CALPHAD models available in FactSage,[23] the thermal conductivity evolution from 273.15 K up to 200K above the melting point of all the eutectic PCMs of interest have been plotted (Figure 2-6). Our calculated thermal conductivity of multicomponent liquid E2 is reasonably reproducing the experimental data reported by Wang et al.[52] Note that even though the composition of E2 is different from the investigated composition of Wang et al.[52], it contains only a small amount of $CaCl_2$ and the compositions of NaCl, KCl and $MgCl_2$ in E2 are similar to those of the PCM reported by Wang et al.[52] In addition, $CaCl_2$ and $MgCl_2$ are chemically close, and their thermal conductivity values of the liquid forms are comparable (Table A2). Hence, we expect the thermal conductivity of the two molten salt mixtures, i.e. E2 and the PCM reported by Wang et al.[52], to be similar as shown in Figure 3. Errors bars for the experimental data are inserted in Figure 3 to show the possible differences of thermal conductivity of the two compositions. The good agreement obtained with the available experimental data of effective thermal conductivity demonstrates the reliable predictive capability of our model.

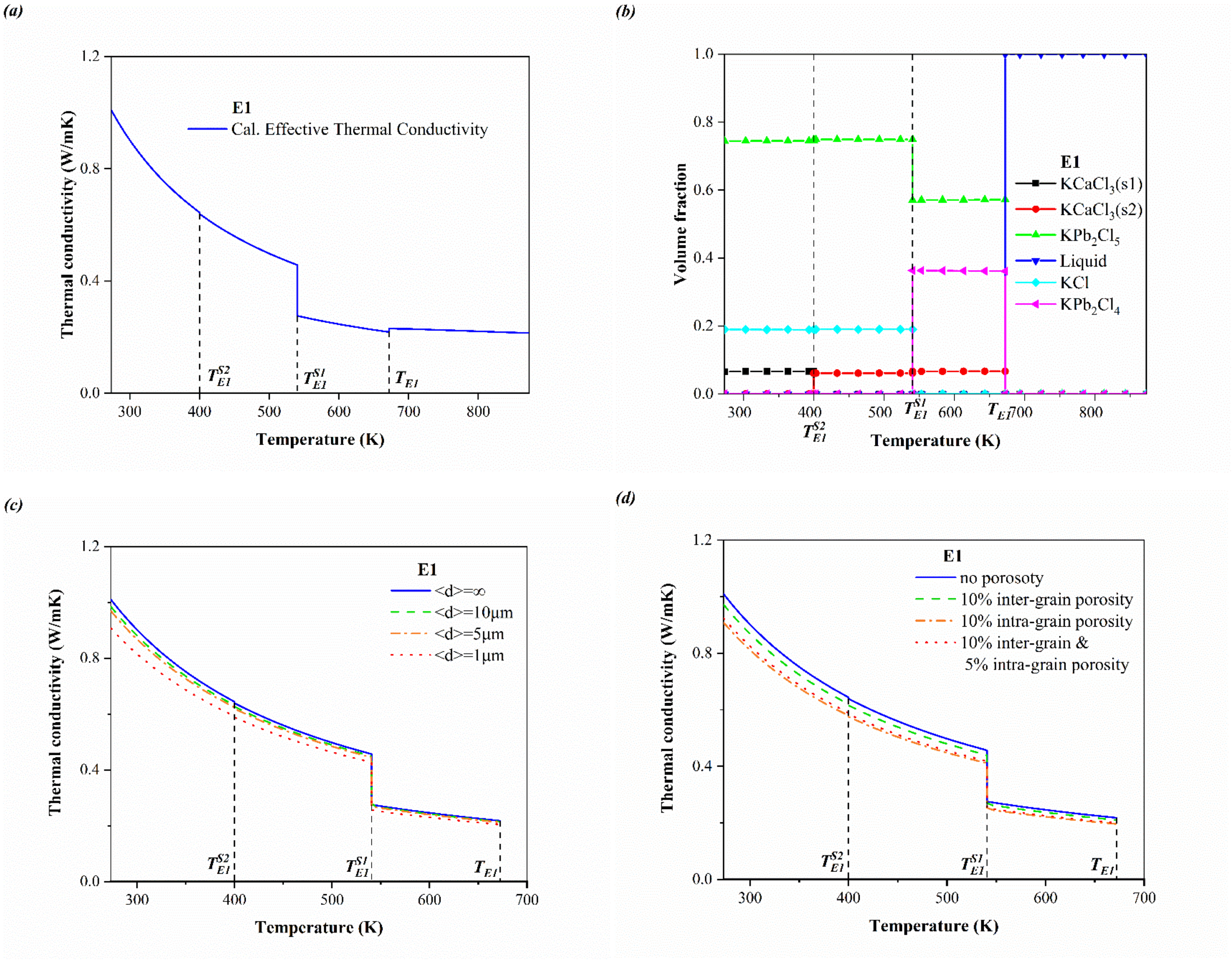

*Figure 2: Evolution of (a) the calculated effective thermal conductivity and (b) the calculated volume fractions of phases with temperature for the E1 PCM. Effect of (c) average grain size and (d) either or both inter-grain porosity and intra-grain porosity on the calculation of the effective thermal conductivity of E1.*

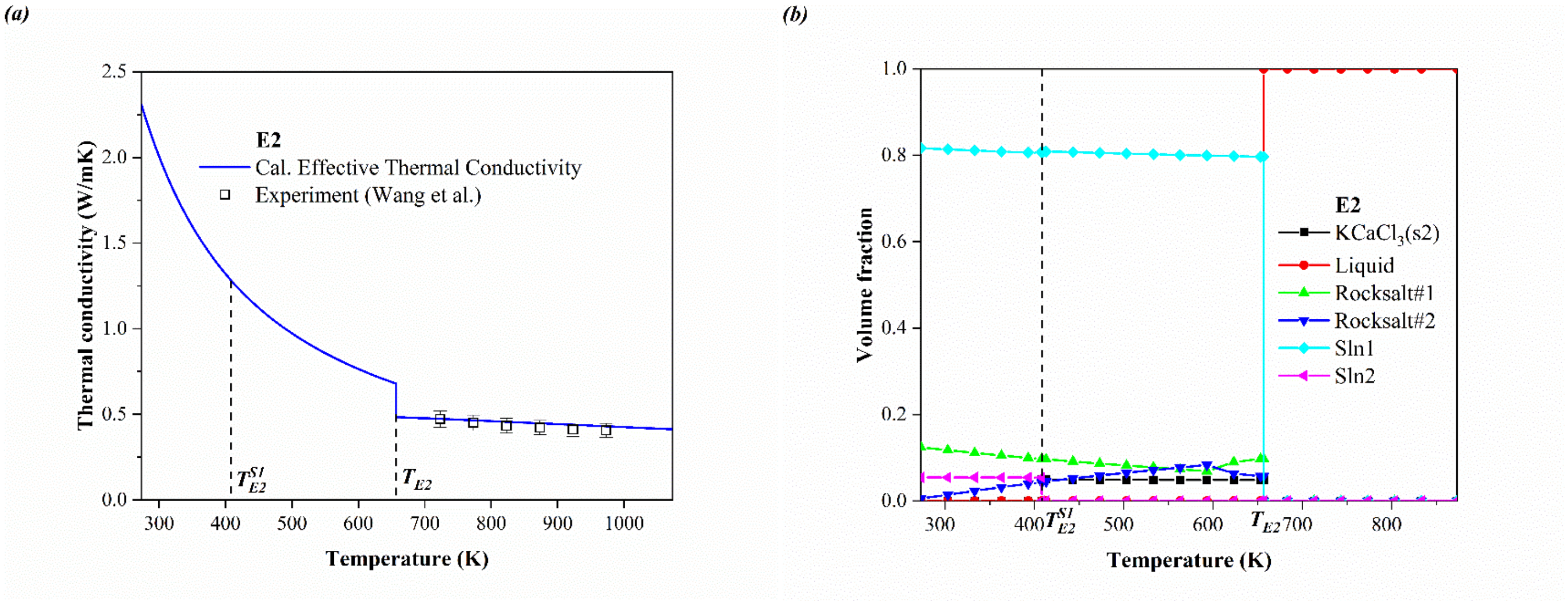

*Figure 3: Evolution of (a) the calculated effective thermal conductivity and (b) the calculated volume fractions of phases with temperature for the E2 PCM. The calculated effective thermal conductivity of liquid E2 is compared with experimental data of thermal conductivity of 0.2046 NaCl – 0.4131 KCl – 0.3823 $MgCl_2$ liquid phase.*[52]

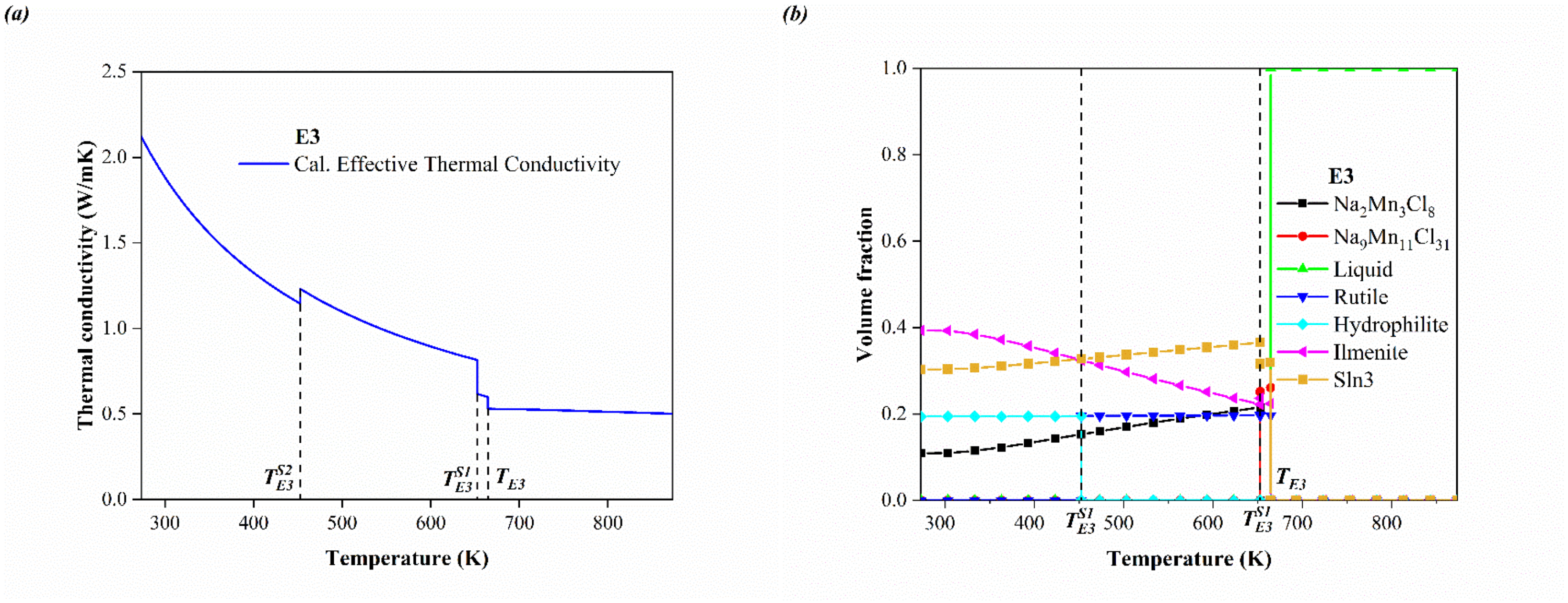

*(c)*

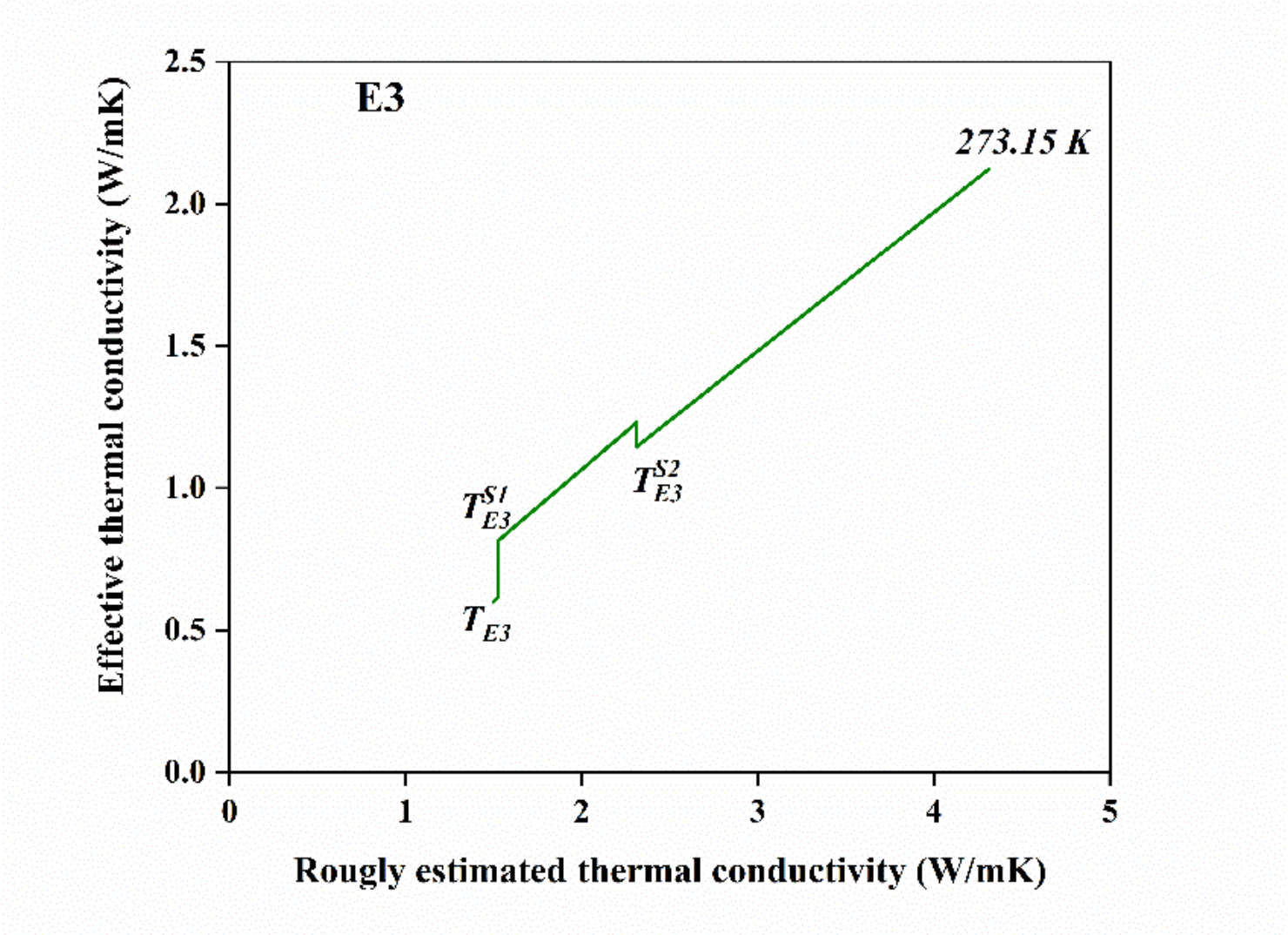

*Figure 4: Evolution of (a) the calculated effective thermal conductivity and (b) the calculated volume fractions of phases with temperature for the E3 PCM . (c) Comparison of our calculated effective thermal conductivity and the roughly estimated thermal conductivity from the composition of E3.*

*(a)* *(b)*

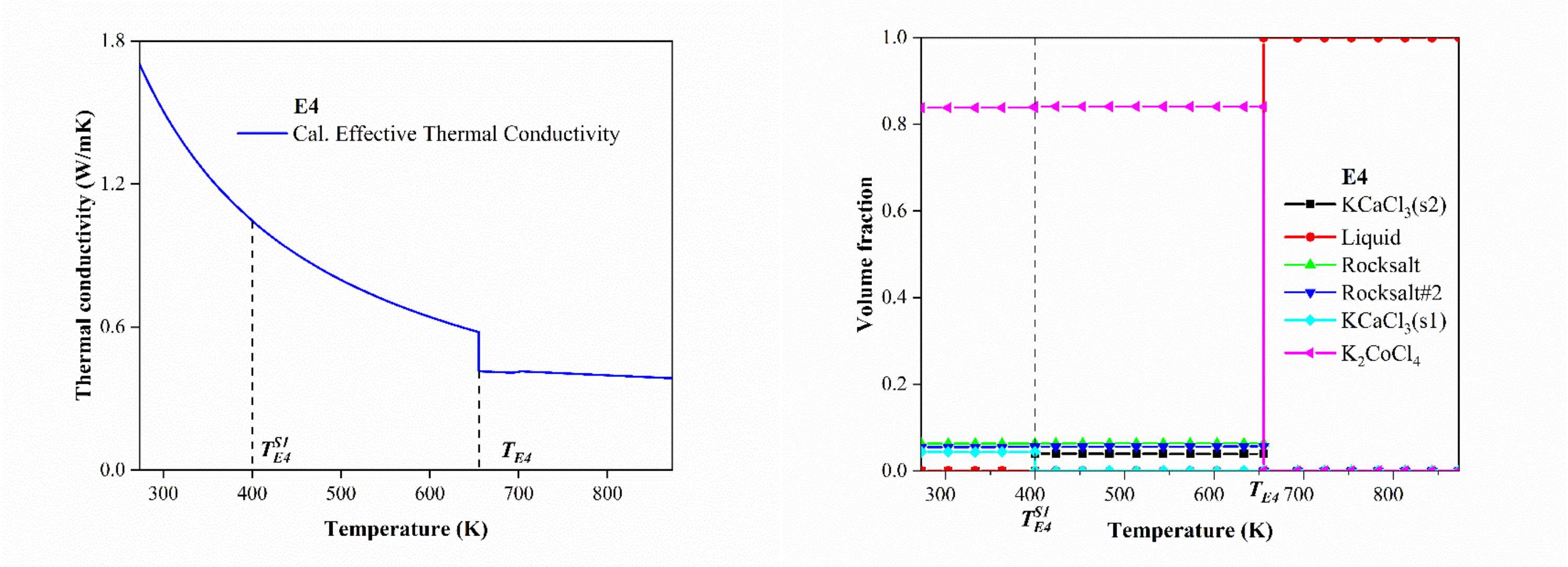

*Figure 5: Evolution of (a) the calculated effective thermal conductivity and (b) the calculated volume fractions of phases with temperature for the E4 PCM.*

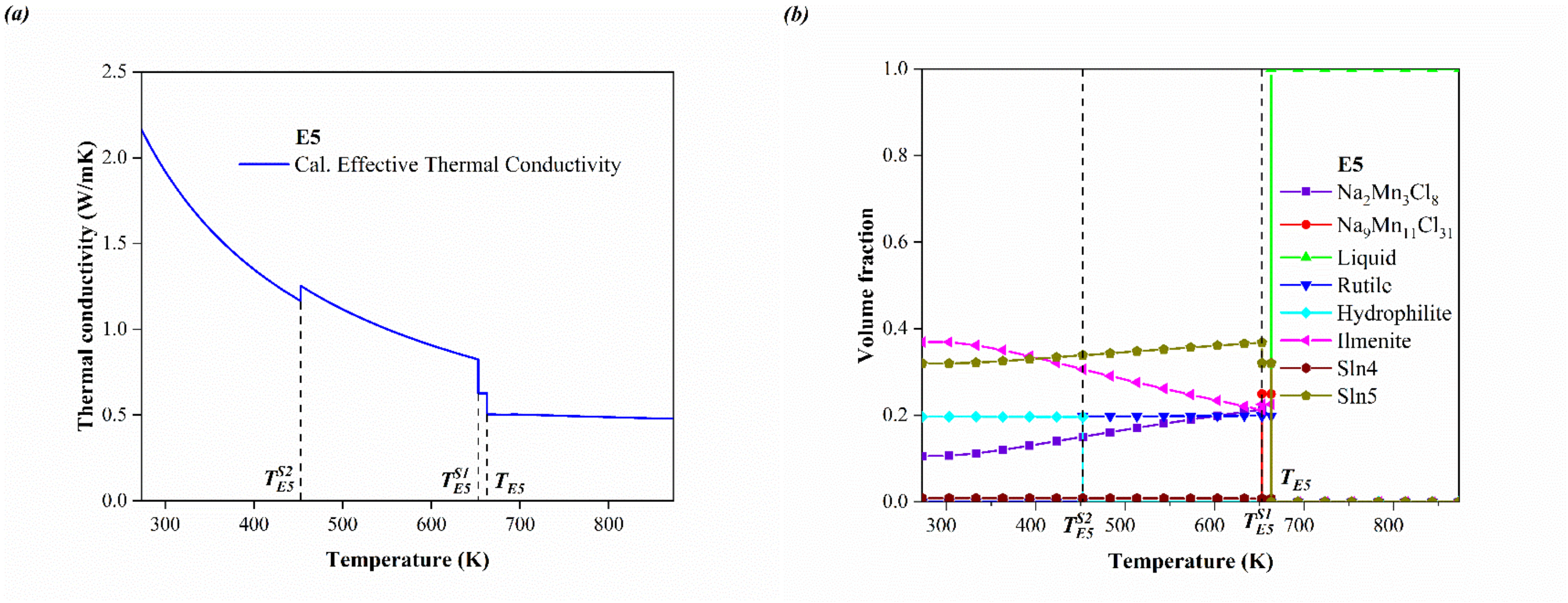

*Figure 6: Evolution of (a) the calculated effective thermal conductivity and (b) the calculated volume fractions of phases with temperature for the E5 PCM.*

Our calculation of effective thermal conductivity also shows that none of the 5 investigated PCMs satisfies the requirement of having a thermal conductivity in solid and liquid states of >1 W/(m.K) which ensures a minimum temperature gradient and consequently accelerates the heat transfer processes[12] (Table 2, Figure 2-6). It is understandable since pure chlorides, in general, have low thermal conductivity (Table A2, A4). In fact, chloride multi-component mixtures are considered promising PCMs because of their high heat of fusion and low cost, despite their low thermal conductivity.[1] Low absolute values of thermal conductivity of pure chlorides generate low values of roughly estimated thermal conductivity, $\lambda^*$, which is calculated directly from volume fraction, $\varphi_i$, and thermal conductivity, $\lambda_i$, of the compound chlorides $i$ forming a PCM ($\lambda^* = \sum \varphi_i \cdot \lambda_i$; e.g. for E3 PCM, $i$= NaCl, $MgCl_2$, $CaCl_2$, $MnCl_2$). It can be seen from our calculations that the effective thermal conductivity of solid state of E3, $\lambda_{eff}^{0}$, is about one half of the roughly estimated one, $\lambda^*$, in a wide temperature range (from 273.15 K up to the eutectic temperature) (Figure 4c). In the case of E3, the solid state at any temperature consists of four different solid solutions and compounds (Table 1). At the same temperature, the thermal conductivity of any of the solid solutions or compounds in the solid state of E3 is normally a few times smaller than that of the simple chlorides forming E3 (i.e. NaCl, $MgCl_2$, $CaCl_2$, and $MnCl_2$) due to their high number of atoms per primitive cell. In general, a solid state of a PCM consisting of several solid solutions and/or complex compounds should have a significantly lower thermal conductivity than the simple compounds forming the PCM. It means that neglecting the phase constitution of solid state and only considering the composition of the material would provide a severely wrong estimation of the thermal conductivity (about 2 times in the case of E3). Therefore, basic knowledge of phase fractions and phase compositions contributing to a solid state is mandatory for an effective prediction of its thermal conductivity.

Besides the low absolute value of thermal conductivity, any hasty alternations of thermal conductivity of a PCM at phase transformation temperatures should be investigated cautiously and monitored with care because it considerably affects the heat transfer process and therefore the operation of the CSP. There exists an immediate increase of the thermal conductivity of every investigated molten salt mixture at the eutectic temperature as the liquid phase transforms into the solid phase assemblage (Figure 3-6). A lower thermal conductivity or less effective thermal energy transport of the liquid phase is associated with a larger intermolecular spacing and more random movements of the molecules in the liquid state compared to the solid state. Furthermore, our calculations also show abrupt changes in the effective thermal conductivity of the solid phase assemblages at the temperatures of solid phase transformations (Figure 2-6). In some cases, like E1, E3, E5, a sudden drop of the effective thermal conductivity within a range of temperature is due to the stability of a complex compound or solid solution comprised of large molecules. For example, the sudden decline of the effective thermal conductivity of E3 at $T_{E3}^{S1}$ is mainly caused by the solid phase transformation from $Na_2Mn_3Cl_8$ which has 13 atoms per cell to $Na_9Mn_{11}Cl_{31}$ which has 51 atoms per cell (Figure 4). The larger the molecule of a complex compound or solid solution, the more unlikely it is to have a high thermal conductivity (Eq. 3), therefore, the formation of any complex compound or solid solution during phase transformations is unfavorable for any promising PCMs.

After the thermal conductivity of the PCM in its solid state has been calculated based on equilibrium properties obtained via FactSage from PCM nominal composition, the effect of non-equilibrium microstructural parameters (i.e. average grain size, porosity) on the calculated effective thermal conductivity is treated. The effects of average grain size and porosity are quantified in our calculations of the effective thermal conductivity of the solid state E1 PCM (Figure 2c, d) by using the Microstructure Module of FactSage 8.1.[23] The effective thermal conductivity of solid state E1 decreases with a decrease of the average grain size and increase of porosity (Figure 2c, d). In general, the thermal conductivity of a polycrystalline is considerably lower than that of the corresponding single crystal due to the thermal resistance at the grain boundaries. The thermal resistance of the grain boundary is a decreasing function of the average grain size (Figure 2c). The increase of grain boundary thermal resistance is commonly attributed to the shortening of the phonon mean free path. The grain size influence is more significant at low temperatures (Figure 2c), where the phonon mean free path is comparable to the average grain size. Due to the boundary effect, the thermal conductivity loss is from 6 to 10% as the average grain size of the E1 solid state reduces to 1 μm (Figure 2c). A loss of about 10% of thermal conductivity of a solid state with the average grain size of 1 μm is commonly reported in various materials such as MgO, α-$Al_2O_3$, $UO_2$, $SnO_2$, $SrTiO_3$, $LiCoO_2$, β-$Si_3N_4$, β-SiC, Si, $Fe_2Sb$...[41, 53-54] In addition, the discontinuation of movements of the molecules in a solid state due to porosity should certainly diminish the effective thermal conductivity (Figure 2d). Intra-grain porosity (nano-porosity) has a stronger effect than inter-grain porosity (macro-porosity). The thermal conductivity declines by only 3.7% with 10% inter-grain porosity while a drop of 10% is observed in the case of 10% intra-grain porosity (Figure 2d). There has been no clear explanation as to why inter-grain porosity has a weaker effect on thermal conductivity than intra-grain porosity. One possible explanation is that as intra-grain porosity occurs within individual grains, it then reduces the phonon mean free path significantly. If inter-grain porosity stayed at the grain boundaries, it would just slightly reduce the phonon mean free path, hence reducing the thermal conductivity less significantly. In summary, both the reduction of the average grain size and augmentation of inter-grain or/and intra-grain porosity result in the interruption of movements of the molecules in a solid state hence diminishing its effective thermal conductivity.

# 4 Conclusion

To sum up, we have combined a series of theoretical models supported by DFT-based atomistic simulations for examining the evolution of the thermal conductivity of multicomponent systems from standard temperature (273.15 K) up to hundreds K above the melting temperature. The models have been applied for 5 promising chloride PCMs which are identified by using the CALPHAD database and MADS algorithm. A good agreement between our calculations and the available experimental data has been revealed, proving the reliability of the model. We have shown that it is critical to have a reliable thermodynamic database correctly estimating the phase fractions and phase constitutions because without it, based only on the nominal composition PCMs, the thermal conductivity can be incorrectly predicted with a substantial error. Besides the expected sudden change of thermal conductivity of the PCMs at the liquid-solid state transformation, there exists rapid alternations of thermal conductivity due to the solid phase transformation, allowing the formation of complex compounds or solid solutions. The more complex compounds or solid solutions with a high number of atoms per primitive cell contributing to the solid state, the lower the value of thermal conductivity. A diminution of thermal conductivity is also attributed to a small average grain size and/or a large inter-grain and/or intra-grain porosity. We believe that the current model of thermal conductivity as a function of temperature is reliable and helpful for engineers and scientists when designing PCMs for improving TES of CSP. To obtain better thermal conductivity, the chosen PCMs should have a maximized amount of simple crystallographic structures (i.e. minimized amount of complex compounds or complex solid solutions), a large average grain size, and a minimal porosity.

## Acknowledgments

We acknowledge the support of the Natural Sciences and Engineering Research Council of Canada (NSERC) [funding reference number: RGPIN-2021-03279]. Computations were made on the supercomputer Bèluga at the Ècole de technologie supérieure, managed by] Calcul-Quèbec and Compute Canada.

## Nomenclature

| | |
|---|---|
| $a$ | Thermal diffusivity |
| $\alpha_m$ | Thermal expansion coefficient |
| $B_s$ | Isentropic bulk modulus |
| $C_{V,m}$ | Heat capacity at constant volume at the melting temperature |
| $C_p$ | Heat capacity at constant pressure |
| $\langle d_i \rangle$ | Average grain size of the phase *i* |
| $\delta^3$ | Volume per atom |
| $\delta_M^\lambda$ | Mass fluctuation effect upon the thermal conductivity |
| $\Gamma$ | Disorder scattering parameter |
| $\Gamma_M$ | Mass fluctuation contribution to the disorder scattering parameter |
| $\Gamma_S$ | Local strain field fluctuation contribution to the disorder scattering parameter |
| $\hbar$ | Planck's constant |
| $k_B$ | Boltzmann's constant |
| $M$ | Average molecule weight of the solution |
| $M_i$ | Molecule weight of constituent $i$ of the solution |
| $\bar{M}$ | Average atomic mass |
| $N_A$ | Avogadro's constant |
| $n$ | Number of atoms per primitive cell |
| $n_0$ | Number of ions per formula of a molten salt compound |
| $\lambda$ | Thermal conductivity |
| $\lambda^*$ | Roughly estimated thermal conductivity of a PCM |
| $\lambda_i$ | Thermal conductivity of the phase *i* or compound *i* |
| $\lambda_{lc}$ | Thermal conductivity of a molten salt compound |
| $\lambda_{ls}$ | Thermal conductivity of a non-reciprocal molten salt mixture |
| $\lambda_\sigma$ | Ideal thermal conductivity of a hypothetical molten mixture |
| $\lambda_{sc}$ | Vibrational (phonon) thermal conductivity of a solid salt compound |
| $\lambda_{ss}$ | Thermal conductivity of a solid solution |
| $\lambda_{ss}^{id}$ | Thermal conductivity of the parent materials |
| $\lambda_{eff}^{ms}$ | Thermal conductivity of the solid system |
| $P_C$ | Critical porosity |
| $P^{inter}$ | Inter-grain porosity |
| $p_i^{intra}$ | Intra-grain porosity (nano-porosity) of the phase *i* |
| $\varphi_i$ | Volume fraction of the phase *i* or compound *i* |
| $r_a$ | Effective anionic radius of a molten salt compound |
| $r_c$ | Effective cationic radius of a molten salt compound |
| $\rho$ | Density |
| $\sigma_i$ | Phonon characteristic length within the grain constituting phase *i* |
| $T$ | Temperature |
| $T_m$ | Melting temperature |
| $\tau$ | Function describing the average inter-grain porosity level upon the entire solid state system |
| $\theta_D$ | Debye temperature |
| $U_m$ | Velocity of sound of a molten salt/mixture |
| $\boldsymbol{u}$ | Disorder scaling parameter within a solid solution |
| $V_m$ | Molar volume |
| $X_i$ | Composition of component *i* |
| $\chi$ | Critical function describing the inter-grain porosity effect |
| $\gamma$ | Grüneisen's constant |

# Appendix

***Model parameters for volumetric thermal expansion coefficient and thermal conductivity of solid and liquid compounds***

*Table A1: Parameters of volumetric thermal expansion coefficient (in the expression* $\boldsymbol{\alpha} = \boldsymbol{a} + \boldsymbol{b} \cdot \boldsymbol{T} + \boldsymbol{c} \cdot \boldsymbol{T}^{-1} + \boldsymbol{d} \cdot \boldsymbol{T}^{-2}$ *) of liquid compounds either taken from Gheribi et al.*[14] *(shown in gray rows) or* [(*)]*calculated in this study from the reported density as a function of temperature*[51] *(shown in orange rows):*

| **Compound** | $\boldsymbol{a} \cdot \mathbf{10^5}$ | $\boldsymbol{b} \cdot \mathbf{10^8}$ | $\boldsymbol{c} \cdot \mathbf{10^2}$ | $\boldsymbol{d}$ |
|---|---|---|---|---|
| **NaCl** | -0.9097 | 21.1815 | 17.4962 | -37.2781 |
| **KCl** | -3.4284 | 25.3743 | 19.6886 | -40.7254 |
| **$CaCl_2$** | 12.0890 | 5.6992 | 2.8884 | -5.9007 |
| **$MgCl_2$** | 10.1371 | 4.0303 | 2.4330 | -4.8892 |
| **$PbCl_2$** | 14.8151 | 13.6598 | 4.8791 | 0.0488 |
| **$MnCl_2$** | 12.9296 | 4.5945 | 1.6596 | -3.1379 |
| **$CoCl_2$**(*) | 20.2680 | 11.9430 | 0.0000 | 0.0000 |
| **$NiCl_2$**(*) | 15.9150 | 6.9568 | 0.0000 | 0.0000 |

*Table A2: Parameters of the expression $\lambda_{lc} = a + b \cdot T$ derived from Eq. 1 for calculating the thermal conductivity of liquid compounds estimated in this study (shown in orange rows), or taken from the literature (shown in gray rows with the notation: (*) for Gheribi et al.[14] and (**) for Gheribi et al.[50]):*

| Compound | ***a*** (Wm$^{-1}$K$^{-1}$) | ***b*** (10$^{-4}$ Wm$^{-1}$K$^{-2}$) |
|---|---|---|
| **NaCl**(*) | 0.718872 | -2.11 |
| **KCl**(*) | 0.606200 | -2.00 |
| **$CaCl_2$**(*) | 0.566975 | -1.15 |
| **$MgCl_2$**(**) | 0.561385 | -0.93 |
| **$PbCl_2$** | 0.350457 | -1.08 |
| **$MnCl_2$** | 0.614910 | -1.24 |
| **$CoCl_2$** | 0.524698 | -1.58 |
| **$NiCl_2$** | 0.637573 | -1.48 |

*Table A3: Our parameters of volumetric thermal expansion coefficient (in the expression $\boldsymbol{\alpha = a + b \cdot T + c \cdot T^{-1} + d \cdot T^{-2}}$) of solid compounds obtained by fitting calculations of the thermodynamically self-consistent (TSC) method and the available experimental data in the literature (note that all the compounds are considered in their stable crystal structures):*

| Compound | Space group | $a \cdot 10^5$ | $b \cdot 10^8$ | $c \cdot 10^2$ | $d$ |
|---|---|---|---|---|---|
| **NaCl** | $Fm\bar{3}m$ | -16.9575 | 27.3690 | 11.6180 | -15.8986 |
| **KCl** | $Fm\bar{3}m$ | -20.1425 | 28.7146 | 12.8700 | -18.0428 |
| **RbCl** | $Fm\bar{3}m$ | 7.5252 | 7.8643 | 1.0310 | -1.5124 |
| **CsCl** | $Pm\bar{3}m$ | 11.2275 | 12.6426 | -0.5160 | 0.5834 |
| **$CaCl_2$** | *Pnnm* | 5.2870 | 0.3237 | 0.0137 | -0.0183 |
| **$SrCl_2$** | $Fm\bar{3}m$ | -19.9326 | 24.9624 | 10.9780 | -15.2230 |
| **$BaCl_2$** | *Pnma* | 5.9813 | 1.1728 | 0.1583 | -0.3444 |
| **$PbCl_2$** | *Pnma* | 6.5776 | 7.4528 | 2.7268 | -4.3904 |
| **$MnCl_2$** | $R\bar{3}m$ | -4.9150 | 37.5794 | 10.9112 | -14.8275 |
| **$MgCl_2$** | $R\bar{3}m$ | 7.8446 | 0.7695 | 0.0439 | -0.0574 |
| **$NiCl_2$** | $R\bar{3}m$ | 8.1654 | 5.2615 | 1.3750 | -2.3700 |
| **$KCaCl_3$** | *Pnma* | 7.6592 | 2.1358 | 0.2633 | -0.6125 |
| **$KCaCl_3$** | *I4/mcm* | 8.1820 | 2.6597 | 0.3590 | -0.7558 |
| **$K_2PbCl_4$** | $I\bar{4}3d$ | -9.4341 | 20.1929 | 8.8059 | -12.575 |
| **$KPb_2Cl_5$** | $P2_1/c$ | 6.0002 | 0.9850 | 0.0563 | -0.1373 |
| **$Na_2Mn_3Cl_8$** | $R\bar{3}m$ | 6.8996 | 1.3464 | 0.0730 | -0.3028 |
| **$Na_9Mn_{11}Cl_{31}$** | $R\bar{3}c$ | 10.5135 | 5.0271 | 0.5624 | -0.9879 |
| **$Na_2MnCl_4$** | *Pbam* | 7.6708 | 1.8123 | 0.1261 | -0.4443 |
| **$NaMgCl_3$** | $R\bar{3}c$ | 8.0373 | 2.5102 | 0.3319 | -0.6399 |
| **$Na_2CoCl_4$** | *Pnma* | 8.1334 | 2.3449 | 0.2422 | -0.5652 |
| **$K_2CoCl_4$** | *Pnma* | 8.7352 | 2.9597 | 0.3404 | -0.6655 |
| **$Na_2NiCl_4$** | *Pbam* | 9.8709 | 4.4371 | 0.5367 | -0.8363 |

*Table A4: Parameter sets of solid compounds used for calculating the thermal conductivity according to Eq. 3 (note that (*)data taken from Gheribi et al.[16] are shown in gray rows, our calculations are in orange rows):*

| **Compound** | $\boldsymbol{\gamma}$ **Grüneisen parameter** | $\boldsymbol{\theta}$ **Debye temperature (K)** | $\boldsymbol{n}$ **number of atoms per primitive cell** | $\boldsymbol{\rho}$ **Density at 298.15 K (g/cm$^3$)** |
|---|---|---|---|---|
| **NaCl(*)** | 1.65 | 285 | 2 | 2.165 |
| **KCl(*)** | 1.41 | 232 | 2 | 1.984 |
| **RbCl** | 1.45 | 145 | 2 | 2.800 |
| **CsCl** | 1.80 | 120 | 2 | 3.988 |
| **$CaCl_2$** | 1.50 | 244 | 6 | 2.174 |
| **$SrCl_2$** | 1.80 | 195 | 3 | 3.052 |
| **$BaCl_2$** | 1.70 | 197 | 24 | 3.856 |
| **$PbCl_2$** | 1.50 | 132 | 12 | 5.550 |
| **$MnCl_2$** | 1.20 | 150 | 3 | 2.977 |
| **$MgCl_2$** | 1.50 | 187 | 3 | 2.319 |
| **$NiCl_2$** | 1.30 | 249 | 3 | 3.530 |
| **$KCaCl_3$**(pnma) | 1.80 | 259 | 20 | 2.140 |
| **$KCaCl_3$** (I4/mcm) | 1.80 | 255 | 20 | 2.340 |
| **$K_2PbCl_4$** | 1.50 | 119 | 7 | 3.65 |
| **$KPb_2Cl_5$** | 1.50 | 155 | 32 | 4.790 |
| **$Na_2Mn_3Cl_8$** | 1.60 | 256 | 13 | 2.734 |
| **$Na_9Mn_{11}Cl_{31}$** | 1.80 | 236 | 51 | 2.698 |
| **$Na_2MnCl_4$** | 1.70 | 281 | 14 | 2.521 |
| **$NaMgCl_3$** | 1.53 | 217 | 10 | 2.258 |
| **$Na_2CoCl_4$** | 1.50 | 251 | 28 | 2.662 |
| **$K_2CoCl_4$** | 1.80 | 279 | 28 | 2.450 |
| **$Na_2NiCl_4$** | 1.47 | 188 | 14 | 2.724 |

# References

1. Kenisarin, M. M., High-temperature phase change materials for thermal energy storage. *Renewable and Sustainable Energy Reviews* **2010,** *14* (3), 955-970.

2. Pielichowska, K.; Pielichowski, K., Phase change materials for thermal energy storage. *Progress in Materials Science* **2014,** *65*, 67-123.

3. Li, M.; Mu, B., Effect of different dimensional carbon materials on the properties and application of phase change materials: A review. *Applied Energy* **2019,** *242*, 695-715.

4. Gulfam, R.; Zhang, P.; Meng, Z., Advanced thermal systems driven by paraffin-based phase change materials – A review. *Applied Energy* **2019,** *238*, 582-611.

5. Du, K.; Calautit, J.; Wang, Z.; Wu, Y.; Liu, H., A review of the applications of phase change materials in cooling, heating and power generation in different temperature ranges. *Applied Energy* **2018,** *220*, 242-273.

6. Bale, C. W.; Bélisle, E.; Chartrand, P.; Decterov, S. A.; Eriksson, G.; Gheribi, A. E.; Hack, K.; Jung, I. H.; Kang, Y. B.; Melançon, J.; Pelton, A. D.; Petersen, S.; Robelin, C.; Sangster, J.; Spencer, P.; Van Ende, M. A., FactSage thermochemical software and databases, 2010–2016. *Calphad* **2016,** *54*, 35-53.

7. Gheribi, A. E.; Robelin, C.; Digabel, S. L.; Audet, C.; Pelton, A. D., Calculating all local minima on liquidus surfaces using the FactSage software and databases and the Mesh Adaptive Direct Search algorithm. *The Journal of Chemical Thermodynamics* **2011,** *43* (9), 1323-1330.

8. Gheribi, A. E.; Audet, C.; Le Digabel, S.; Bélisle, E.; Bale, C. W.; Pelton, A. D., Calculating optimal conditions for alloy and process design using thermodynamic and property databases, the FactSage software and the Mesh Adaptive Direct Search algorithm. *Calphad* **2012,** *36*, 135-143.

9. Gheribi, A. E.; Harvey, J.-P.; Bélisle, E.; Robelin, C.; Chartrand, P.; Pelton, A. D.; Bale, C. W.; Le Digabel, S., Use of a biobjective direct search algorithm in the process design of material science applications. *Optimization and Engineering* **2016,** *17* (1), 27-45.

10. Gheribi, A. E.; Le Digabel, S.; Audet, C.; Chartrand, P., Identifying optimal conditions for magnesium based alloy design using the Mesh Adaptive Direct Search algorithm. *Thermochimica Acta* **2013,** *559*, 107-110.

11. Gheribi, A. E.; Pelton, A. D.; Harvey, J.-P., Determination of optimal compositions and properties for phase change materials in a solar electric generating station. *Solar Energy Materials and Solar Cells* **2020,** *210*, 110506.

12. Zhou, C.; Wu, S., Medium- and high-temperature latent heat thermal energy storage: Material database, system review, and corrosivity assessment. *International Journal of Energy Research* **2019,** *43* (2), 621-661.

13. Zhao, J.; Ma, T.; Li, Z.; Song, A., Year-round performance analysis of a photovoltaic panel coupled with phase change material. *Applied Energy* **2019,** *245*, 51-64.

14. Gheribi, A. E.; Torres, J. A.; Chartrand, P., Recommended values for the thermal conductivity of molten salts between the melting and boiling points. *Solar Energy Materials and Solar Cells* **2014,** *126*, 11-25.

15. Gheribi, A. E.; Chartrand, P., Application of the CALPHAD method to predict the thermal conductivity in dielectric and semiconductor crystals. *Calphad* **2012,** *39*, 70-79.

16. Gheribi, A. E.; Poncsák, S.; St-Pierre, R.; Kiss, L. I.; Chartrand, P., Thermal conductivity of halide solid solutions: Measurement and prediction. *The Journal of Chemical Physics* **2014,** *141* (10), 104508.

17. Gheribi, A. E.; Poncsák, S.; Guérard, S.; Bilodeau, J.-F.; Kiss, L.; Chartrand, P., Thermal conductivity of the sideledge in aluminium electrolysis cells: Experiments and numerical modelling. *The Journal of Chemical Physics* **2017,** *146* (11), 114701.

18. Gheribi, A. E.; Salanne, M.; Chartrand, P., Formulation of Temperature-Dependent Thermal Conductivity of NaF, β-Na3AlF6, Na5Al3F14, and Molten Na3AlF6 Supported by Equilibrium Molecular Dynamics and Density Functional Theory. *The Journal of Physical Chemistry C* **2016,** *120* (40), 22873-22886.

19. Gheribi, A. E.; Chartrand, P., Thermal Conductivity of Compounds Present in the Side Ledge in Aluminium Electrolysis Cells. *JOM* **2017,** *69* (11), 2412-2417.

20. Gheribi, A. E.; Chartrand, P., Thermal conductivity of molten salt mixtures: Theoretical model supported by equilibrium molecular dynamics simulations. *The Journal of Chemical Physics* **2016,** *144* (8), 084506.

21. Kohler, F., Zur Berechnung der thermodynamischen Daten eines ternären Systems aus den zugehörigen binären Systemen. *Monatshefte für Chemie und verwandte Teile anderer Wissenschaften* **1960,** *91* (4), 738-740.

22. Toop, G. W., Predicting Ternary Activities Using Binary Data. *Trans. TMS-AIME* **1965,** *223*, 850-855.

23. https://www.factsage.com/.

24. Renaud, E.; Robelin, C.; Gheribi, A. E.; Chartrand, P., Thermodynamic evaluation and optimization of the Li, Na, K, Mg, Ca, Sr // F, Cl reciprocal system. *The Journal of Chemical Thermodynamics* **2011,** *43* (8), 1286-1298.

25. Bukhalova, G. A.; Mirsoyanova, N. N.; Yagub; x; yan, E. S., Theoretical and experimental investigations of the quinary mutual system of fluorides and chlorides of lithium, sodium, potassium and strontium. **1981**.

26. Morelli, D. T.; Slack, G. A., High Lattice Thermal Conductivity Solids. In *High Thermal Conductivity Materials*, Shindé, S. L.; Goela, J. S., Eds. Springer New York: New York, NY, 2006; pp 37-68.

27. Abeles, B., Lattice Thermal Conductivity of Disordered Semiconductor Alloys at High Temperatures. *Physical Review* **1963,** *131* (5), 1906-1911.

28. Fancher, D. L.; Barsch, G. R., Lattice theory of alkali halide solid solutions—III. Pressure dependence of solid solubility and spinodal decomposition. *Journal of Physics and Chemistry of Solids* **1971,** *32* (6), 1303-1313.

29. Seifitokaldani, A.; Gheribi, A. E., Thermodynamically self-consistent method to predict thermophysical properties of ionic oxides. *Computational Materials Science* **2015,** *108*, 17-26.

30. Seifitokaldani, A.; Gheribi, A. E.; Dollé, M.; Chartrand, P., Thermophysical properties of titanium and vanadium nitrides: Thermodynamically self-consistent approach coupled with density functional theory. *Journal of Alloys and Compounds* **2016,** *662*, 240-251.

31. Gheribi, A. E.; Seifitokaldani, A.; Wu, P.; Chartrand, P., An ab initio method for the prediction of the lattice thermal transport properties of oxide systems: Case study of Li2O and K2O. *Journal of Applied Physics* **2015,** *118* (14), 145101.

32. Saxena, S. C.; Kachhava, C. M., A simple method for calculating thermal expansion of crystals. *Applied Scientific Research* **1966,** *16* (1), 162-166.

33. Srivastava, K. K.; Merchant, H. D., Thermal expansion of alkali halides above 300°K. *Journal of Physics and Chemistry of Solids* **1973,** *34* (12), 2069-2073.

34. Rapp, J. E.; Merchant, H. D., Thermal expansion of alkali halides from 70 to 570 K. *Journal of Applied Physics* **1973,** *44* (9), 3919-3923.

35. He, C.; Hu, C.-E.; Zhang, T.; Qi, Y.-Y.; Chen, X.-R., Lattice dynamics and thermal conductivity of cesium chloride via first-principles investigation. *Solid State Communications* **2017,** *254*, 31-36.

36. Sist, M.; Fischer, K. F. F.; Kasai, H.; Iversen, B. B., Low-Temperature Anharmonicity in Cesium Chloride (CsCl). *Angewandte Chemie International Edition* **2017,** *56* (13), 3625-3629.

37. Gerlich, D.; Andersson, P., Temperature and pressure effects on the thermal conductivity and heat capacity of CsCl, CsBr and CsI. *Journal of Physics C: Solid State Physics* **1982,** *15* (25), 5211-5222.

38. Yu, X.; Hofmeister, A. M., Thermal diffusivity of alkali and silver halide crystals as a function of temperature. *Journal of Applied Physics* **2011,** *109* (3), 033516.

39. Choy, T. C., *Effective Medium Theory: Principles and Applications*. 2 ed.; Oxford University Press: Oxford, 2015; p 240.

40. Landauer, R., Electrical conductivity in inhomogeneous media. *AIP Conference Proceedings* **1978,** *40* (1), 2-45.

41. Gheribi, A. E.; Chartrand, P., Effect of Grain Boundaries on the Lattice Thermal Transport Properties of Insulating Materials: A Predictive Model. *Journal of the American Ceramic Society* **2015,** *98* (3), 888-897.

42. Rhee, S. K., Porosity—Thermal conductivity correlations for ceramic materials. *Materials Science and Engineering* **1975,** *20*, 89-93.

43. Aivazov, M. I.; Domashnev, I. A., Dependence of electrical and thermal conductivities of hot-pressed titanium nitride samples on porosity. *Poroshkovaya Metallurgiya (Kiev)* **1968,** *8*, 51.

44. Gheribi, A. E.; Gardarein, J.-L.; Rigollet, F.; Chartrand, P., Evidence of second order transition induced by the porosity in the thermal conductivity of sintered metals. *APL Materials* **2014,** *2* (7), 076105.

45. Gheribi, A. E.; Gardarein, J.-L.; Autissier, E.; Rigollet, F.; Richou, M.; Chartrand, P., Experimental study of the thermal conductivity of sintered tungsten: Evidence of a critical behaviour with porosity. *Applied Physics Letters* **2015,** *107* (9), 094102.

46. Gheribi, A. E.; Autissier, E.; Gardarein, J.-L.; Richou, M., Thermal transport properties of multiphase sintered metals microstructures. The copper-tungsten system: Experiments and modeling. *Journal of Applied Physics* **2016,** *119* (14), 145104.

47. Chatterjee, A.; Verma, R.; Umashankar, H.; Kasthurirengan, S.; Shivaprakash, N.; Behera, U., Heat conduction model based on percolation theory for thermal conductivity of composites with high volume fraction of filler in base matrix. **2018**.

48. Marcus, Y., The compressibility of molten salts. *The Journal of Chemical Thermodynamics* **2013,** *61*, 7-10.

49. Sternberg, S.; Vasilescu, V., Ultrasonic velocity compressibility, and excess volume for binary mixtures of molten salts MgCl2 + BaCl2, MgCl2 + KCl. *The Journal of Chemical Thermodynamics* **1969,** *1* (6), 595-606.

50. Gheribi, A. E.; Phan, A. T.; Chartrand, P., A theoretical framework for reliable predictions of thermal conductivity of multicomponent molten salts mixtures: KCl-NaCl-MgCl2 as a case study. *Solar Energy Materials and Solar Cells* **2021,** *Submitted*.

51. Janz, G. J.; Dampier, F. W.; Lakshminarayanan, G. R.; Standards, U. S. N. B. o.; Lorenz, P. K., *Molten Salts: Volume 1. Electrical Conductance, Density, and Viscosity Data*. U.S. Government Printing Office: 1968.

52. Wang, X.; Rincon, J. D.; Li, P.; Zhao, Y.; Vidal, J., Thermophysical Properties Experimentally Tested for NaCl-KCl-MgCl2 Eutectic Molten Salt as a Next-Generation High-Temperature Heat Transfer Fluids in Concentrated Solar Power Systems. *Journal of Solar Energy Engineering* **2021,** *143* (4).

53. Shrestha, K.; Yao, T.; Lian, J.; Antonio, D.; Sessim, M.; Tonks, M. R.; Gofryk, K., The grain-size effect on thermal conductivity of uranium dioxide. *Journal of Applied Physics* **2019,** *126* (12), 125116.

54. He, J.; Zhang, L.; Liu, L., Thermal transport in monocrystalline and polycrystalline lithium cobalt oxide. *Physical Chemistry Chemical Physics* **2019,** *21* (23), 12192-12200.